\documentclass[cits]{PoS}
\usepackage{epsfig}
\usepackage{amsfonts}
\usepackage{amsmath}

\title{Finite-size scaling of heavy-light mesons}

\ShortTitle{FSS of hl mesons}

\author{\speaker{Fabio Bernardoni}, Pilar Hern\'andez
\\
          IFIC - Instituto de F\'isica Corpuscular\\
	Edificio Institutos de Investigaci\'on\\
	Apartado de Correos 22085\\
	E-46071 Valencia - Espa\~na \\
         E-mail: \email{fabiob@ific.uv.es, pilar@ific.uv.es}}

\author{Silvia Necco\\
        CERN - PH-TH Division\\
        CH-1211 Geneve 23\\
        Switzerland\\
        E-mail: \email{Silvia.Necco@cern.ch}}

\abstract{We study the finite-size scaling of heavy-light mesons in the static limit.
The most relevant effects are due to the pseudo-Goldstone boson cloud. In the
HMChPT framework we compute two-point functions of left current densitities
as well as pseudoscalar densitites for the cases in which some or all of
them lay in the epsilon-regime. As expected, finite volume dependence turns
out to be significant in this regime and can be predicted in the effective
theory in terms of the infinite-volume low-energy couplings. These results
might be relevant for extraction of heavy-light meson properties from lattice
simulations.
\vspace{2cm}
\begin{flushright}
FTUV-09-1014 \\
IFIC/09-43\\
CERN-PH-TH-2009-181 \\
\end{flushright}

}

\FullConference{The XXVII International Symposium on Lattice Field Theory - LAT2009\\
		 July 26-31 2009\\
		 Peking University, Beijing, China}

\newcommand{\tr}{{\rm Tr\,}}

\newcommand{\ba}{\begin{eqnarray}}
\newcommand{\ea}{\end{eqnarray}}
\newcommand{\be}{\begin{equation}}
\newcommand{\ee}{\end{equation}}
\renewcommand{\vec}[1]{{\bf #1}}

\begin{document}

\section{Introduction}
The simulations of heavy-light (hl) mesons made out of a heavy quark (charm or bottom) and a light one (up, down or strange) on the lattice are challenging because they require very large volumes and small lattice spacings in order to keep systematic errors under control.\\
Requiring that $a \lesssim 1/(2m_b)$ and $M_{ll} L \gtrsim 4$, where $M_{ll}$ is the mass of a pseudo-Goldstone boson made of two light quarks, would imply a simulation at $a \lesssim 0.02$ fm and $L \gtrsim 6$ fm if $M_{ll}\sim M_\pi$, with an $L/a \gtrsim 300$.\\
If the heavy quark mass is sufficiently large a good effective description is provided by heavy quark effective theory (HQET) 
\cite{Grinstein:1990mj,Eichten:1989zv,Georgi:1990um}, which is a systematic expansion in the limit of infinite heavy quark mass. In this limit, the scale $M_{hl}$ (the mass of a heavy-light meson) disappears from the problem and the UV cutoff can in principle be as low as the cutoff used to describe light meson dynamics. Indeed this approximation has been extensively used to simulate heavy-light  mesons in lattice QCD (see \cite{Gamiz:2008iv} for a review on heavy flavour phenomenology on Lattice QCD). \\
On the other side, both if the heavy quark is treated in the static limit or not, an obvious question is whether we can do better concerning the constraint on the box-size. After all the finite-size scaling of heavy-light systems should be dominated by light pions physics, since these are the lightest modes in QCD. To the extent that pion physics can be described by chiral perturbation theory (ChPT), it is conceivable that finite-size scaling of heavy-light systems can be accurately predicted using ChPT, as the finite-size scaling of light mesons is \cite{Gasser:1986vb, Gasser:1987ah}. \\
Having integrated the bottom out, we can take the lattice spacing as big as $a\lesssim 1/(2m_c)\sim0.05$ fm; then in a lattice of length $L = 2 fm$  we can reasonably expect that ChPT still provides a good description of QCD. At this volume, a lattice computation with masses in the epsilon regime
would require $L/a \gtrsim 40$.

In \cite{us} we have computed the finite-size scaling (FSS) of heavy-light systems when the lightest pions are light compared to the inverse box size in two limiting situations. 
In the first one the heavy quark is significantly heavier than the light one, but still treatable in ChPT: this correspond to considering hl mesons in the mixed-regime introduced in \cite{mixed};
in the second one the heavy quark is static and therefore chiral dynamics can be treated in Heavy Meson Chiral Perturbation Theory (HMChPT): this correspond to considering hl static mesons in the $\epsilon$-regime.  \\
Even though these two situations are physically very different, the pion dynamics responsible 
for the finite-size scaling properties should be pretty much the same. It is therefore interesting to see explicitly how a quantitative matching of the finite-size effects takes place, by comparing the finite volume dependence of correlation functions in ChPT and HMChPT. \\
We present the result obtained for the left-handed current correlator to next-to-leading order in both effective theories and discuss its applicability to extract LECs from lattice simulations. Full results and details are presented in \cite{us}.

\section{Heavy-light mesons in the mixed-regime of ChPT}
\label{sec:mixed}

We consider $N_h$ heavy quarks of mass $m_h$ and $N_l$ light quarks of mass $m_l$ with $m_l \ll m_h$, but such that both can still be treated in the context of ChPT. In practice we apply the power counting: 
\begin{equation}
m_l \sim \epsilon^4, \quad m_h \sim \epsilon^2, \quad L^{Ð1} \sim p \sim \epsilon \,,
\label{eq:pc} 
\end{equation}
that has been named {\it mixed-regime} in \cite{mixed}, to the common ChPT Lagrangian.
While we refer to \cite{mixed} for further details on the implementation of ChPT in the mixed-regime, both in the full and partially-quenched theories, we just remind that in this regime the zero modes of the light pions have to be treated non perturbatively. 
Following the same method we have computed the two-point correlation functions of two heavy-light left-handed currents to relative ${\mathcal O}(\epsilon^2)$ order:
\begin{eqnarray}
  \tr[T^aT^b]{C_J}(t) & \equiv & 
 \int \! {\rm d}^3 x\, 
 \Bigl\langle {J}^a_0(x) {J}^b_0(0) \Bigr\rangle  
 \end{eqnarray}
where $ J^a_\mu \equiv \bar\psi T^a \gamma_\mu P_- \psi $ and $T^a$ is a traceless generator of $SU(N_l+N_h)$ with one index in the light and one in the heavy subsector. \\

\subsection{Results in the mixed regime of ChPT}

The result is:
\begin{multline}
\label{eq:caj}
C_J(t)= \frac{F_{(A)}^2}{2}M_{(A)}^2P(t,M_{(A)}^2)-\frac{T}{2V}\Bigg\{ \left(N_l-\frac{1}{N_l}  \right){\overline{k}_{00}(M_{h}^2,t)}\\
+\left(N_h-\frac{1}{N_h}\right) {k_{00}(M_{h}^2,2M_{h}^2,t)}+ \left(\frac{1}{N_h}+\frac{1}{N_l} \right)  {k_{00}(M_{h}^2,M_{\eta}^2,t)}  \Bigg\},
\end{multline}
where we have defined the renormalized $F_{(A)}$ and $M_{(A)}^2$ as:
\begin{eqnarray}
F_{(A)}^2 &=& F^2-\frac{1}{2}\left( N_h-\frac{1}{N_h}  \right)G\left(0,2M_{h}^2\right)-\frac{N_l}{2}\overline{G}\left(0,0\right)-\frac{N_l+N_h}{2}G\left(0,M_{h}^2\right) \label{FANLO}\nonumber \\
&&-\left(\frac{1}{2(N_l+N_h)}+\frac{1}{2N_h} \right)G\left(0,M_{\eta}^2\right)+\frac{1}{2}E^{\epsilon}(0)+8M_{h}^2(2 L_4N_h+L_5) \\
M_{(A)}^2 &=&M_h^2\left[1-\frac{1}{F^2}\left(8M_h^2(2L_4N_h+L_5-4N_hL_6-2L_8) -\frac{2N_h+3N_l}{3(N_l+N_h)^2}G(0,M_{\eta}^2)\right. \right. \nonumber \\
&&\left.\left.+\frac{1}{6}\frac{\Box E^{\epsilon}(0)}{M_h^2}  \right) -\frac{2}{\mu_h}\left(\frac{1}{6N_l}-\frac{N_l}{4}-\frac{\mu_{l}}{4}\langle (U_0+U_0^{\dagger})_{l l} \rangle  \right) \right] \,\,,\label{MANLO}
\end{eqnarray}
and we have defined the shorthands $\mu_i \equiv m_i\Sigma V$, $M_h^2=\frac{\Sigma m_h}{F^2}$ and $M_{\eta}^2=\frac{2N_l}{N_l+N_h}M_{h}^2$. Here we have used:
\ba
\label{def_G}
G(x, M^2) =  \frac{1}{V}\sum_{p}\frac{e^{ipx}}{p^2+M^2}\,, &\quad& \overline{G}(x,M^2) =  \frac{1}{V}\sum_{p\ne 0}\frac{e^{ipx}}{p^2+M^2}\,,  \\
E^\epsilon(x)  =  \frac{1}{V}\sum_{p\neq 0}\frac{e^{ipx}}{(p^2)^2F(p)}-\frac{N_h}{2N_l^2VM_{h}^2}\,, &\quad& 
F(p)=\frac{N_h}{p^2+2M_{h}^2}+\frac{N_l}{p^2}\,.
\ea
The time dependence is expressed in terms of the function:
\be\label{def_P}
P(t, M^2) \equiv \int d^3\vec{x}G(x, M^2) ={\cosh \left[ M \left( {T \over 2} - |t| \right) \right] \over 2 M \sinh\left[ { M T \over 2}\right]}
\ee
and its derivatives:
\be
{k_{00}(M_1^2,M_2^2,t)} \equiv
\frac{1}{2}\sum_{\vec{p}}\Bigg\{2\frac{dP}{dt}(t,M_{1\vec{p}}^2)\frac{dP}{dt}(t,M_{2\vec{p}}^2) -\left(P(t,M_{1\vec{p}}^2)\frac{d^2P}{dt^2}(t,M_{2\vec{p}}^2) + (M_1 \leftrightarrow M_2) \right) \Bigg\}, \\    
\ee
where $M_{a\vec{p}}\equiv \sqrt{M_a^2+\vec{p}^2}$. In this regime we also need:
\be
{\overline{k}_{00}(M_1^2,t)} \equiv  \lim_{M_2\rightarrow 0} \left(k_{00}(M_1^2,M_2^2,t)+\frac{P(t,M_1^2)M_1^2}{2TM_2^2}\right)\,.
\ee

The result in Eq.~(\ref{eq:caj}) represents the FSS of kaon-like states ($m_h=m_s$ and $m_l =m_u=m_d$)  for various situations:
\begin{itemize}
\item $2+1$ dynamical simulations:  $N_h=1, N_l= 2$; 
\item partially quenched (PQ) simulations where the $h$ quarks are quenched and the $l$ quarks are dynamical by taking the replica limit $N_hÊ\rightarrow 0$ of Eq.~(\ref{eq:caj});
\item PQ simulations where the $l$ quarks are all quenched  or PQ, while the $h$ quarks are dynamical. In this case, the appropriate value of $N_l$ must be taken, but also the zero-mode integrals  $\langle (U_0+U_0^{\dagger})_{ll}\rangle$ need to be properly defined \cite{Kanzieper:2002ix,Splittorff:2002eb} \footnote{Note that one cannot consider a fully quenched theory with $N_h=N_l=0$ on the basis of Eq.~(\ref{eq:caj}), because the singlet has been integrated out\cite{mixed2}.}.
\end{itemize}

\section{Heavy-light mesons in the HMChPT}
\label{sec:HMint}

\begin{figure}[htbp]
\begin{center}
$$
\begin{array}{cc}
\includegraphics[width=6cm]{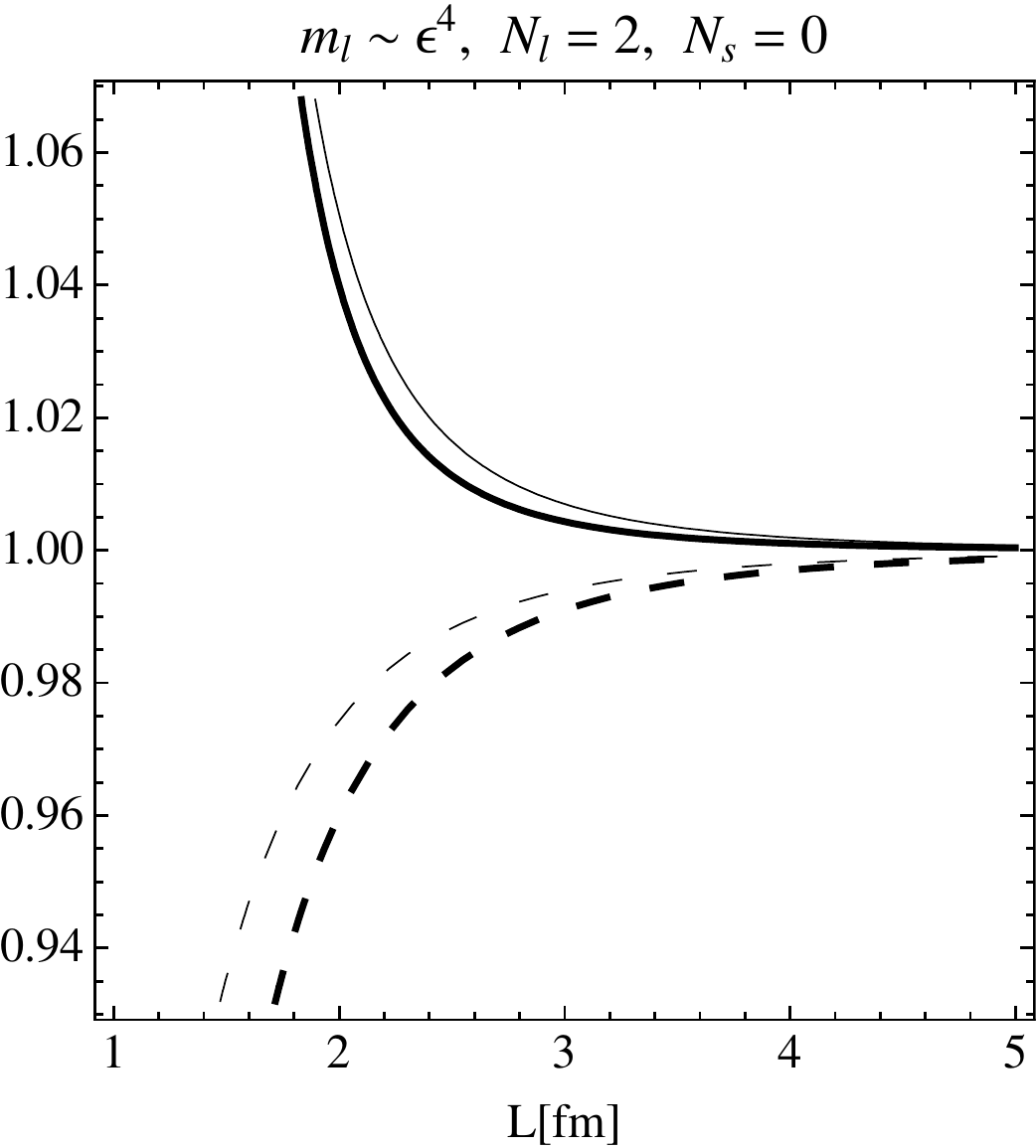} \hspace{0.5cm}&\hspace{0.5cm}\includegraphics[width=6cm]{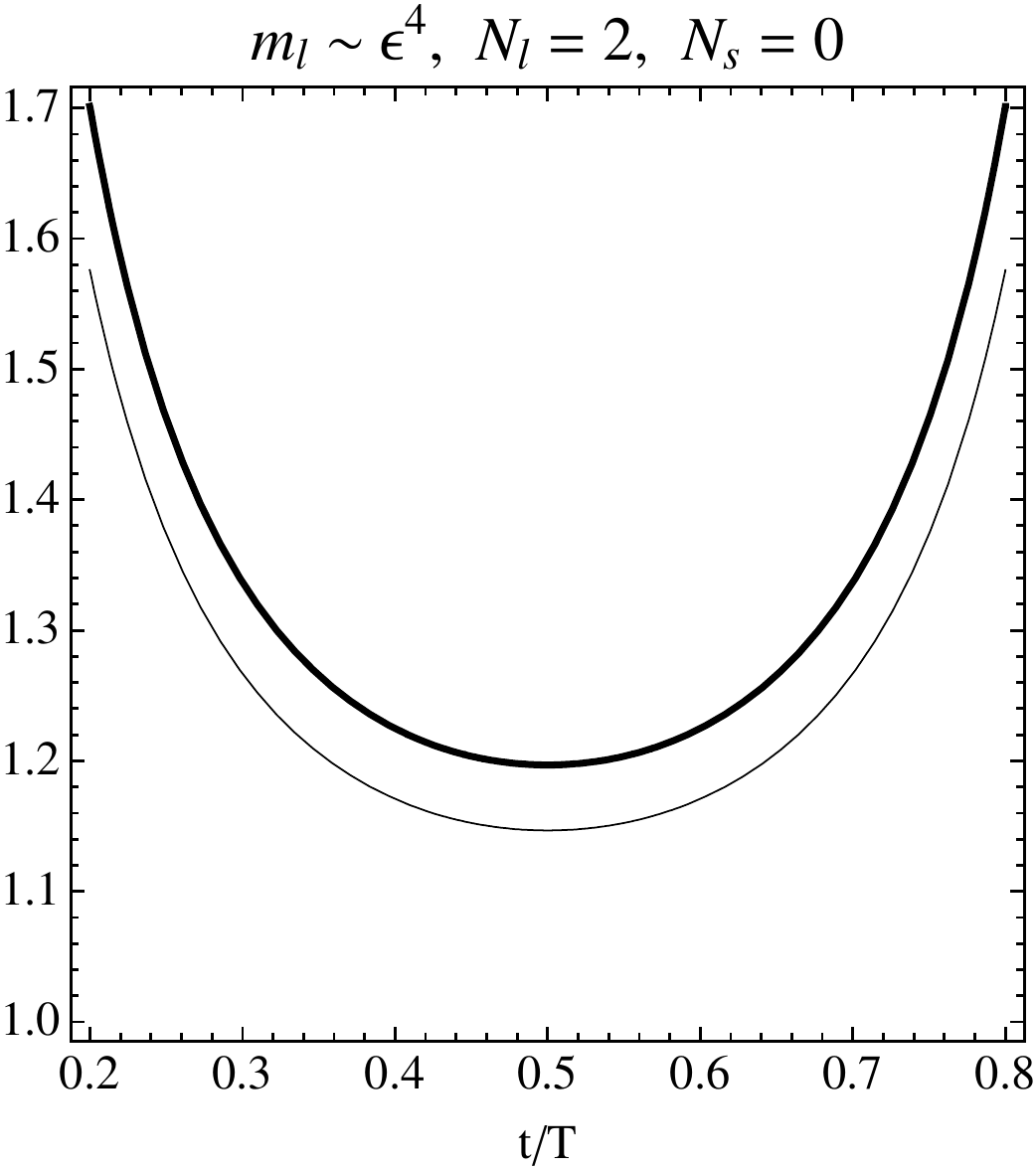} 
\end{array}
$$
\caption{ 
Left: ratio of ${\cal C}_J(t=1 fm)$ at fixed volume normalized to the $\infty$ volume result as a function of $L$ for two lattices with $T=L$ (solid) and $T=2 L$ (dashed), and for $g_\pi=0.44$ (thick lines)\cite{Becirevic:2009yb} and $g_\pi=0$ (thin lines) and typical values of $\Sigma=(250\mbox{MeV})^3, F=90\mbox{MeV}$. Right: $4 {\cal C}_J(t)/(a^2 \exp(-\Delta M^{\epsilon} t))$ as a function of $t/T$  for $T=L=2$ fm, and for $g_\pi=0.44$ (thick line) and $g_\pi=0$ (thin line). }
\label{fig:epsilon}
\end{center}
\end{figure}

The effects of pion dynamics in the properties of static heavy-light mesons can be predicted in 
HMChPT \cite{Burdman:1992gh,Wise:1992hn,Yan:1992gz}. In \cite{Arndt:2004bg}, chiral corrections to B parameters in the $p$-regime were studied. As far as we know, the $\epsilon$-regime has not been explored yet. However the technology we have used was developed in \cite{Smigielski:2007pe} to perform $\epsilon$-regime calculations in baryon ChPT.\\
The most relevant complication is that in the HMChPT Lagrangian half integer powers of the pseudo-Goldstone field appear, so that the observables are not just functional derivatives of the zero modes partition functional (which can be calculated explicitely). We dealt with this problem by working with an explicitely parametrization for the Goldstone bosons manifold. We restricted to the $N_l=2$ case.
The result can be written in the form \footnote{Note that we use ${\cal C}_J (t)$ for HMChPT and $C_J(t)$ for ChPT. The reason for this will be clear in section \ref{sec:lat}. }:
\begin{equation}
{\cal C}_{J}(t)|_{N_l=2} =\frac{a^2 \theta(t)}{8}  \exp\left(-\Delta M^{(\epsilon)} t\right)
\Bigg[1+\frac{3}{4} \frac{1}{(F L)^2 } \left(H(t, L, T) +g_\pi^2 H'(t,L,T) \right) \Bigg]\,,
\label{eq:hmchpt}
\end{equation}
where 
\ba
H(t, L, T) &\equiv & L^2 \left(\frac{T}{L^3}h_1\left(\frac{t}{T}  \right)+\frac{1}{L^3}\sum_{\vec{p}\neq 0}P(t,\vec{p}^2)-\overline{G}(0,0)\right)\,, \quad h_1\left(\frac{t}{T} \right) =\frac{1}{2}\left[\left(\frac{|t|}{T}-\frac{1}{2} \right)^2-\frac{1}{12} \right]\,,\nonumber\\
H'(t, L, T)&\equiv & {1 \over L}  \sum_{\vec{p}\neq 0}
\left(P(t,\vec{p}^2)- P(0, \vec{p}^2)\right)\,,
\label{eq:H}
\ea
and 
\ba
\label{dmasse}
\Delta M^{(\epsilon)} \equiv    \frac{3g_\pi^2}{8 F^2 L^3}\,, \qquad a=F_P\sqrt{2M_P}=\sqrt{2}F_{P^*}/\sqrt{M_{P^*}}\,.
\ea
The coupling $g_\pi$ is responsible for the scattering processes $BB^*\pi$ and $B^*B^*\pi$. It has been computed on the lattice by several authors; the most recent determination has been done in  \cite{Becirevic:2009yb} where the value $g_\pi= 0.44$ has been found. It is interesting to stress that the dependence on $m_l$ and on the topological sector disappears after all diagrams are summed. \\
This result may be used to analyze the scaling for correlators of B or D systems. To analyze the $B_s$ and the $B_s^*$, or to take into account the effects due to the strange quark explicitely for B or $B^*$ systems, one has to add a quark in the p-regime, as we discuss in \cite{us}. However in these systems the influence of the scalar and axial resonances is not negligible, as argued in \cite{Becirevic:2007dg}, and the validity of HMChPT without including them is questionable.\\
In Fig.~\ref{fig:epsilon} (left), we show the ratio of the finite-volume to infinite volume correlator in Eq.~(\ref{eq:hmchpt}) at $t=1$ fm as a function of L for two different geometries and two values of $g_\pi$. Corrections are ${\cal O}(4\%)$ at 2 fm, and the depedence on $g_\pi$ is mild.
In Fig.~\ref{fig:epsilon} (right), we show the time dependence of Eq.~(\ref{eq:hmchpt}) after factoring out the $\exp(- \Delta M^{(\epsilon)} t)$ (we will see later that in any real fit, $\Delta M$ would renormalize the hl meson mass).  

\section{Matching}

Dominant finite-size effects in QCD are due to pion dynamics, since these are the lightest degrees of freedom. It is therefore expected that the finite-size scaling of heavy-light systems does not 
depend on the large energy scales related to the heavy quarks, ie. $M_{h}$ or $M_{\eta}$. This must be the case as long as those scales are significantly larger than $L^{-1}$. Whether these scales are much larger also than the QCD scale so that the static limit (HQET) is a good approximation, or not, 
should not matter a priori for the finite-size scaling properties, because the volume dependence arises from the propagation of  the light degrees of freedom. \\
The leading finite volume effects are therefore expected to come from the fact that the heavy meson can emit and absorb a pion. The probability for this to happen can however depend on the heavy mass scale. 
Close enough to the chiral limit, the masses of pseudoscalar mesons are suppressed by the spontaneous breaking of chiral symmetry, so, for example, we do not need to include the vector mesons in the effective theory, because they are much heavier and decouple.  To the contrary, in the limit $m_h \to \infty$ pseudoscalar and vector mesons are degenerate because the interaction between quark and antiquark inside the meson becomes spin independent, so they both need to be considered in HMChPT. The presence of heavy-light vector resonances can modify the finite volume effects indirectly by inducing unsuppressed  contributions to pion/heavy-light meson scattering. We will see that indeed the finite-size corrections in HMChPT and mixed ChPT match up to corrections proportional to $g^2_\pi$. 

We show now how the matching works. Given any meson two-point function, the first point to realize is that a finite static limit is recovered after factorizing out the leading $e^{- Mt}$, where $M$ is the mass of the hl meson. What can be matched is the dependence of the correlators on the volume, that is $L$ and $T$ and the masses of the up, down and strange quarks because these are explicit degrees of freedom in both effective theories. Moreover, since we only consider the static limit of HMChPT, we have to drop from the ChPT result those contributions that are suppressed by negative powers of $m_h$. In order to match eqs.~(\ref{eq:hmchpt}) and (\ref{eq:caj}),  the L, T dependence must be the same in both cases. 
For the mixed ChPT framework, we split the contribution due to the heavy pions from the rest in Eqs. (\ref{FANLO}) and (\ref{MANLO}) and write:
\ba
F_{(A)}^2&=&  \overline{F}^2(m_h,\,N_l)- {1 \over 2} \left(N_l - {1\over N_l}\right)\overline{G}(0,0)+O(m_h^{-1})\,,\\
M_{(A)}^2&=& \overline{M}_h^2(m_h,\,N_l)+O(m_h^{-1})\,\,.
\ea
$ \overline{F}$ and $\overline{M}_h$ have absorbed the dependence on the heavy quark mass. These expressions just remind us that $F$ and $\Sigma$ depend on the number of quarks that are considered in the effective theory. 
The static limit of the mixed ChPT case in Eq.~(\ref{eq:caj}) is, for $t>0$:
\begin{eqnarray}
{C_{J}(t)\over \exp(-\overline{M}_ht)} &=& {\overline{F}^2\overline{M}_h\over 4} \left[ 1   + {1 \over 2 F^2L^2} \left(N_l - {1\over N_l}\right) H(t,L,T) \right], 
\label{eq:deg}
\end{eqnarray}
where $H(t,L,T)$ is the function of Eq.~(\ref{eq:H}).

For $N_l=2$, the result is identical to the NLO prediction ${\cal C}_{J}(t)$ in HMChPT (Eq.~(\ref{eq:hmchpt})) with the following identifications:
\begin{eqnarray}
a=F_P\sqrt{2M_P}=\overline{F}\sqrt{2\overline{M}_h}, \qquad  g_\pi = 0 .
\label{identifications}
\end{eqnarray}
The fact that at NLO we have to put $g_{\pi}=0$ to match the two expressions reflects the fact that the vector meson is integrated out in the chiral theory. 

\section{Finite-size scaling of heavy-light mesons in lattice QCD}
\label{sec:lat}

As we have seen above the matching of finite-size effects of heavy-light correlators in HMChPT and ChPT works as expected. We would be interested however to use these results to predict the finite-size scaling of these correlators  computed in lattice QCD. We can do this including a relativistic heavy quark or in the static limit. In both cases we expect that for sufficiently large time separation:
\ba
C^{lat}_J(t) \equiv \sum_{\vec{x}} \langle J_\mu^a(x) J_\mu^a(0) \rangle_{lat} \simeq  {\cal C}_J(t) \times {1\over 2M}\exp(- M t),
\label{eq:genbeh}
\ea
where $M$ is the lightest heavy-light meson mass $M_{hl}$ in the case of a relativistic heavy quark or the so-called static energy, $E_{stat} = M_{hl}-m_h$ in the lattice static limit.  \\
Note that the value of $E_{stat}$ is not predicted by HMChPT, however in general we can write:
\be
E_{stat} =E_{stat}^{(0)}+\Delta M \,,
\ee
where $E_{stat}^{(0)}$ is the value the static energy would have in the chiral limit, while $\Delta M$ contains the chiral corrections that are predicted by HMChPT. In the $\epsilon$-regime case they are given in Eq.~(\ref{dmasse}). \\
In practice this means that to fit a correlator using Eq.~ (\ref{eq:hmchpt}) one has to determine four parameters: $a$, $F$, $E_{stat}^{(0)}$ and $g_\pi$. It remains to be seen what the stability of such fits is in practice. 


\end{document}